\pdfoutput=1

\documentclass[sigconf]{acmart}

\copyrightyear{2017} 
\acmYear{2017} 
\setcopyright{acmlicensed}
\acmConference{CIKM'17 }{November 6--10, 2017}{Singapore, Singapore}\acmPrice{15.00}\acmDOI{10.1145/3132847.3133015}
\acmISBN{978-1-4503-4918-5/17/11}

\usepackage{booktabs, threeparttable, tabularx, multirow}  
\usepackage{graphicx, subcaption, xcolor}  
\usepackage{amsmath, amssymb, amsfonts, bm}   
\usepackage{algorithm, algpseudocode}   
\usepackage{natbib}
\usepackage{underscore}

\floatname{algorithm}{Procedure}

\setcitestyle{square,comma,numbers,sort&compress,sectionbib}

\begin{document}

\title{Active Sampling for Large-scale Information Retrieval Evaluation}

\author{Dan Li}
\affiliation{%
  \institution{University of Amsterdam}
  \city{Amsterdam, The Netherlands} 
}
\email{d.li@uva.nl}

\author{Evangelos Kanoulas}
\affiliation{%
  \institution{University of Amsterdam}
  \city{Amsterdam, The Netherlands} 
}
\email{e.kanoulas@uva.nl}

\begin{abstract}
  Evaluation is crucial in Information Retrieval. The development of models, tools and methods has significantly benefited from the availability of reusable test collections formed through a standardized and thoroughly tested methodology, known as the {\em Cranfield paradigm}. Constructing these collections requires obtaining relevance judgments for a {\em pool} of documents, retrieved by systems participating in an evaluation task; thus involves immense human labor. To alleviate this effort different methods for constructing collections have been proposed in the literature, falling under two broad categories: (a) sampling, and (b) active selection of documents. The former devises a smart sampling strategy by choosing only a subset of documents to be assessed and inferring evaluation measure on the basis of the obtained sample; the sampling distribution is being fixed at the beginning of the process. The latter recognizes that systems contributing documents to be judged vary in quality, and actively selects documents from good systems. The quality of systems is measured every time a new document is being judged. In this paper we seek to solve the problem of large-scale retrieval evaluation combining the two approaches. We devise an active sampling method that avoids the bias of the active selection methods towards good systems, and at the same time reduces the variance of the current sampling approaches by placing a distribution over systems, which varies as judgments become available. We validate the proposed method using TREC data and demonstrate the advantages of this new method compared to past approaches.
\end{abstract}



\maketitle

\section{Introduction}

Evaluation is crucial in Information Retrieval (IR). The development of models, tools and methods has significantly benefited from the availability of reusable test collections formed through a standardized and thoroughly tested methodology, known as the Cranfield paradigm~\cite{cleverdon1967cranfield}. Under the Cranfield paradigm the evaluation of retrieval systems typically involves assembling a document collection, creating a set of information needs (topics), and identifying a set of documents relevant to the topics. 

One of the simplifying assumptions made by the Cranfield paradigm is that the relevance judgments are complete, i.e. for each topic all relevant documents in the collection have been identified. When the document collection is large identifying all relevant documents is difficult due to the immense human labor required. In order to avoid judging the entire document collection {\em depth-k pooling}~\cite{Jones:1975} is being used: a set of retrieval systems (also called {\em runs}) ranks the document collection against each topic, and only the union of the top-$k$ retrieved documents is being assessed by human assessors. Documents outside the depth-k pool are considered irrelevant. Pooling aims at being fair to all runs and hopes for a diverse set of submitted runs that can provide a good coverage of all relevant documents. Nevertheless, the underestimation of recall~\cite{Zobel:1998} and the pooling bias generated when re-using these pooled collections to evaluate novel systems that retrieve relevant but unjudged documents~\cite{Zobel:1998, Buckley:2007, Webber:2009, Lipani:2016a} are well-known problems.

The literature suggests a number of approaches to cope with missing judgments (an overview can be found in \cite{Sanderson:2010} and \cite{Kanoulas:2015}): (1) Defining IR measures that are robust to missing judgments, like bpref~\cite{Buckley:2004}. The developed measures however may not precisely capture the notion of retrieval effectiveness one requires, while some have been shown to remain biased~\cite{Yilmaz:2006}. (2) Running a meta-experiment where runs are ``left out'' from contributing to the pool and measuring the bias experienced by these left-out runs compared to the original pool, which is then used to correct measurements over new retrieval systems~\cite{Webber:2009, Lipani:2016a, Lipani:2016b, Lipani:2016c}. (3) Leaving the design of the evaluation measure unrestricted, but instead introducing a document selection methodology that carefully chooses which documents to be judged. Methods proposed under this approach belong to two categories: (a) sample-based methods~\cite{aslam2006statistical,Yilmaz:2006,pavlu2007practical,Yilmaz:2008, Schnabel:2016}, and (b) active selection methods~\cite{cormack1998efficient, Aslam:2003, losada2016feeling, lipani2017fixed}.

Sample-based methods devise a sampling strategy that randomly selects a subset of documents to be assessed; evaluation measures are then inferred on the basis of the obtained sample. Different methods employ different sampling distributions. \citet{aslam2006statistical} and \citet{Yilmaz:2006} use a uniform distribution over the ranked document collection, while \citet{pavlu2007practical} and \citet{Yilmaz:2008} recognize that relevant documents typically reside at the top of the ranked lists returned by participating runs and use stratified sampling to draw larger sample from the top ranks. \citet{Schnabel:2016} also use a weighted-importance sampling method on documents with the sampling distribution optimized for a comparative evaluation between runs. In all aforementioned work, an experiment that dictates the probability distribution under which documents are being sampled is being designed in such a way that evaluation measures can be defined as the expected outcome of this experiment. Evaluation measures can then be estimated by the judged documents sampled. In all cases the sampling distribution is being defined at the beginning of the sampling process and remains fixed throughout the experiment. Sample-based methods have the following desirable properties: (1) on average, estimates have no systematic error, (2) past data can be re-used by new, novel runs without introducing bias, and (3) sampling distributions can be designed to optimize the number of judgments needed to confidently and accurately estimate a measure.

On the other hand, active-selection methods recognize that systems contributing documents to the pool vary in quality. Based on this observation they bias the selection of documents towards those retrieved by good retrieval systems. The selection process is deterministic and depends on how accurately the methods can estimate the quality of each retrieval system. Judging is performed in multiple rounds: at each round the best system is identified, and the next unjudged document in the ranked list of this system is selected to be judged. The quality of systems is calculated at the end of each round, as soon as a new judgment becomes available. Active-selection methods include Move-to-Front~\cite{cormack1998efficient}, Fixed-Budget Pooling~\cite{lipani2017fixed}, and Multi-Armed Bandits~\cite{losada2016feeling}. \citet{losada2016feeling} considers the problem as an exploration-exploitation dilemma, balancing between selecting documents from the best-quality run, and exploring the possibility that the quality of some runs might be underestimated at different rounds of the experiment. The advantage of active-selection methods compared to sample-based methods is that they are designed to identify as many relevant document as possible, by selecting documents with the highest relevance probability. The disadvantage is that the judging process is not fair to all runs, with the selection of documents being biased towards good-performing runs.

 
In this paper, we follow a sample-based approach for an efficient large-scale evaluation. Different from past sample-based approaches we account for the fact that some systems are of higher quality than others, and we design our sampling distribution to over-sample documents from these systems. At the same time, given that our approach is a sample-based approach the estimated evaluation measures are, by construction, unbiased on average, and judgments can be used to evaluate new, novel systems without introducing any systematic error. The method we propose therefore is an {\em active sampling} method with the probability distribution over documents changing at every round of judgments through the re-estimation of the quality of the participating runs. Accordingly, our solution consists of a \textit{sampling} step and an \textit{estimation} step. In the sampling step, we construct a distribution over runs and a distribution over documents in a ranked list and calculate a joint distribution over documents to sample from. In the estimation step, we use the Horvitz-Thompson estimator to correct for the bias in the sampling distribution and estimate evaluation measure values for all the runs. The estimated measures then dictate the new sampling distribution over systems, and hence a new joint distribution over the ranked collection of documents.

Therefore, the contribution of this paper is a new sampling methodology for large-scale retrieval evaluation that combines the advantages of the sample-based and the active-selection approaches. We demonstrate that the proposed method outperforms state-of-the-art methods in terms of effectiveness, efficiency, and reusability.

\section{Active sampling}

In this section we introduce our new sampling method.


\begin{table}[ht]
\centering
\caption{Notation used throughout this paper}
\begin{tabular}{l  l }
\toprule
Symbol         & Description     \\ \hline
$\mathcal{C}$  & $Depth-k$ document collection    \\
$S$            & Sample set \\
$S'$           & Subset of $S$, only containing unique documents \\
$N$            & Total number of unique documents in $\mathcal{C}$  \\
$K$            & Total number of contributing runs \\
$T$            & Number of sampling rounds    \\
$N_b$          & Number of unique documents sampled in round $t$ \\
$N_t$          & Number of documents sampled in round $t$ \\
$d_i$          & $i$-th document \\
$y_i$          & Relevance of document $d_i$ \\
$r(i)$         & Rank of document $d_i$ \\
$p_{t}(k)$    & Probability of $k$-th system run being sampled \\
\multirow{ 2}{*}{$p_{t}(k, r(i))$}    &  Probability of the document ranked $r$ in  $k$-th \\
			& system run  being sampled  \\

\bottomrule
\end{tabular}
\label{tbl:Notation}
\end{table}

\subsection{Active sampling algorithm}
\begin{figure}[h!]
\centering
\includegraphics[width = 0.5\textwidth]{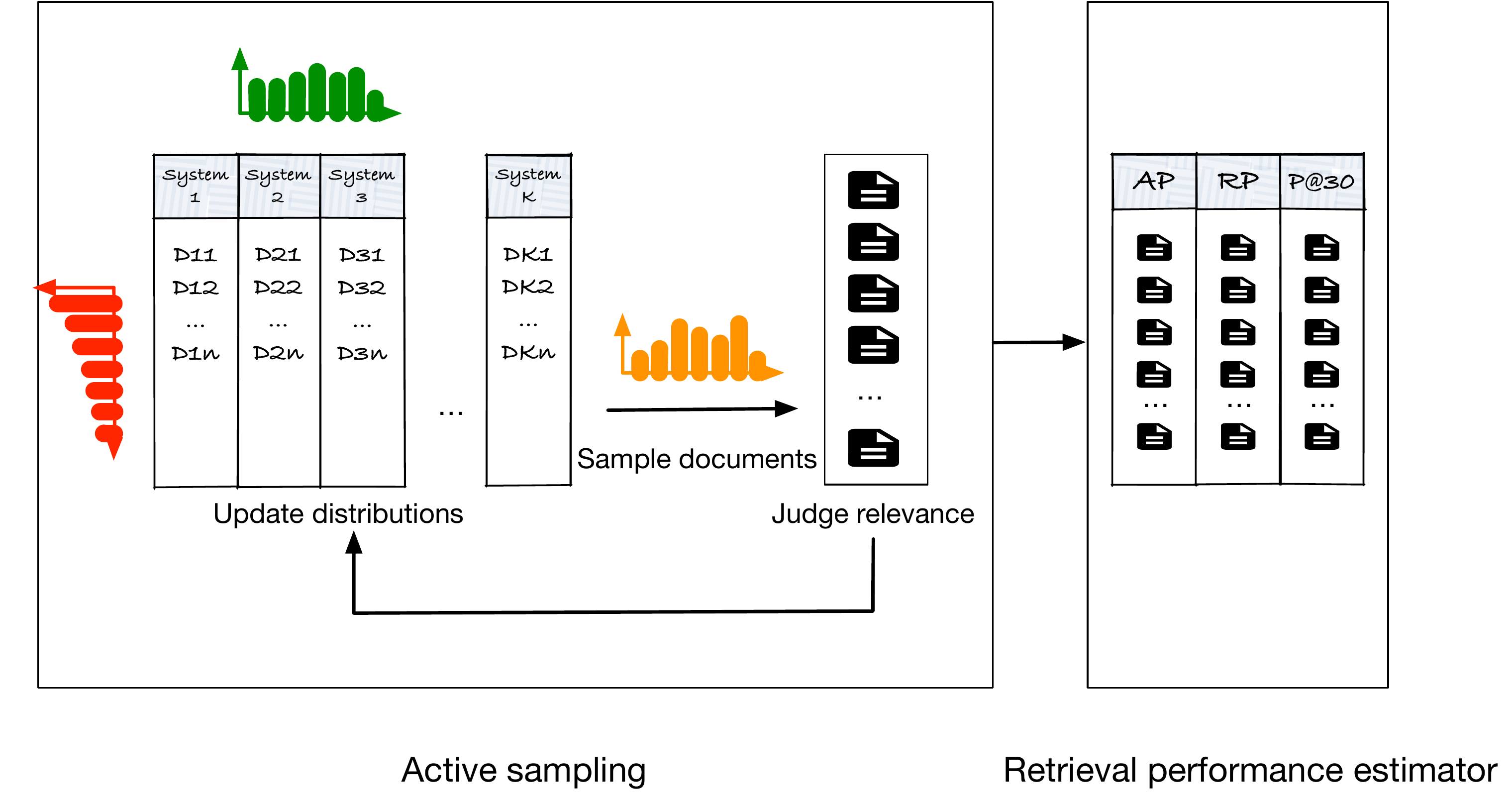}
\caption{Active sampling and retrieval performance estimation}
\label{fig:ApproachFig}
\end{figure}

The key idea underlying our sampling strategy is to place a probability distribution over runs and a probability distribution over documents in the ranked lists of the runs, and iteratively sample documents from the joint distribution. At each round, we sample a set of documents from the joint probability distribution (batch sampling) and request relevance judgments by human assessors. The judged documents are then used to update the probability distribution over runs. The process is repeated until we reach a fixed budget of human assessments (Figure~\ref{fig:ApproachFig}).

\begin{algorithm*}
\caption{Active sampling}\label{alg:sampling}
\begin{algorithmic}[1]
\Require Prior distribution over runs $\big \{p_1(k) \big \}_{k=1}^{K}$, prior distributions over document ranks $\big \{ p_{1}(k, r(i)) \big \} _{i=1}^{N}$, document collection $\mathcal{C}$, batch size $N_b$
\Ensure Sampled documents $S$, associated with relevance judgment and selection probability: $ \big \{ (d_{t,j}, y_{t,j}, p_t(j)) \big \}_{j=1}^{N_bT}$
\For {$t = 1, 2, ..., T $}
\State Calculate the joint document sampling distribution $p_t(i) = \sum_{k=1}^{K}{ p_{t}(k) p_{t}(k, r(i)) , i=1, ..., N}$
\State Sample $N_t$  documents with replacement  (so that it contains $N_b$ unique documents) from $p_t(i)$ 
\State Let the sampled document be $d_{t, j}$; judge relevance of the sampled documents $y_{t, j}, j=1,...,N_t$
\State Augment data $S_{t+1} = S_{t} \bigcup \big \{ (d_{t,j}, y_{t,j}) \big \}_{j=1}^{N_{t}} $
\State Update distribution over runs $p_{t+1}(k), k=1,...,K $
\EndFor
\end{algorithmic}
\end{algorithm*}
 
The process is illustrated in Algorithm \ref{alg:sampling}, while Table 1 shows the notation used throughout the paper. Initially, we provide a prior distribution over runs $\big \{p_1(k)\big \}_{k=1}^{K}$, a prior distribution over the ranks of the documents $\big \{ p_1(k, r(i)) \big \}_{i=1}^{N}$ for each run $k$ , and the document collection $\mathcal{C}$. Given that we have no prior knowledge of the system quality it is reasonable to use a uniform probability distribution over runs, i.e. $p_1 (1) = p_1 (2) = ... = p_1 (K)$. At each round $t$, we calculate the selection probabilities of the documents (that is the probability that a document is selected at each sampling time) $p_t(i)$ for each document $i$, and then sample a document on the basis of this distribution. We use \textit{sampling with replacement with varying probabilities} to sample documents, which is closely related to how we calculate the unbiased estimators and it is describe in Section 3. The sampled documents $d_{t, j}$ are then judged by human assessors, with the relevance of these documents denoted as $y_{t, j}$, and the new data are added to $S_{t}$ which is used to update the ${(t+1)}^{th}$ posterior distribution over runs.


\subsection{Distribution over runs}
The distribution over runs determines the probability of sampling documents from each run. Similar to active-selection methods, we make the assumption that good systems retrieve more relevant documents at the top of their rankings compared to bad systems. Based on this assumption we wish to over-sample from rankings of good systems. 


%
%

\begin{figure}[h!]
\centering
\includegraphics[width = .4\textwidth]{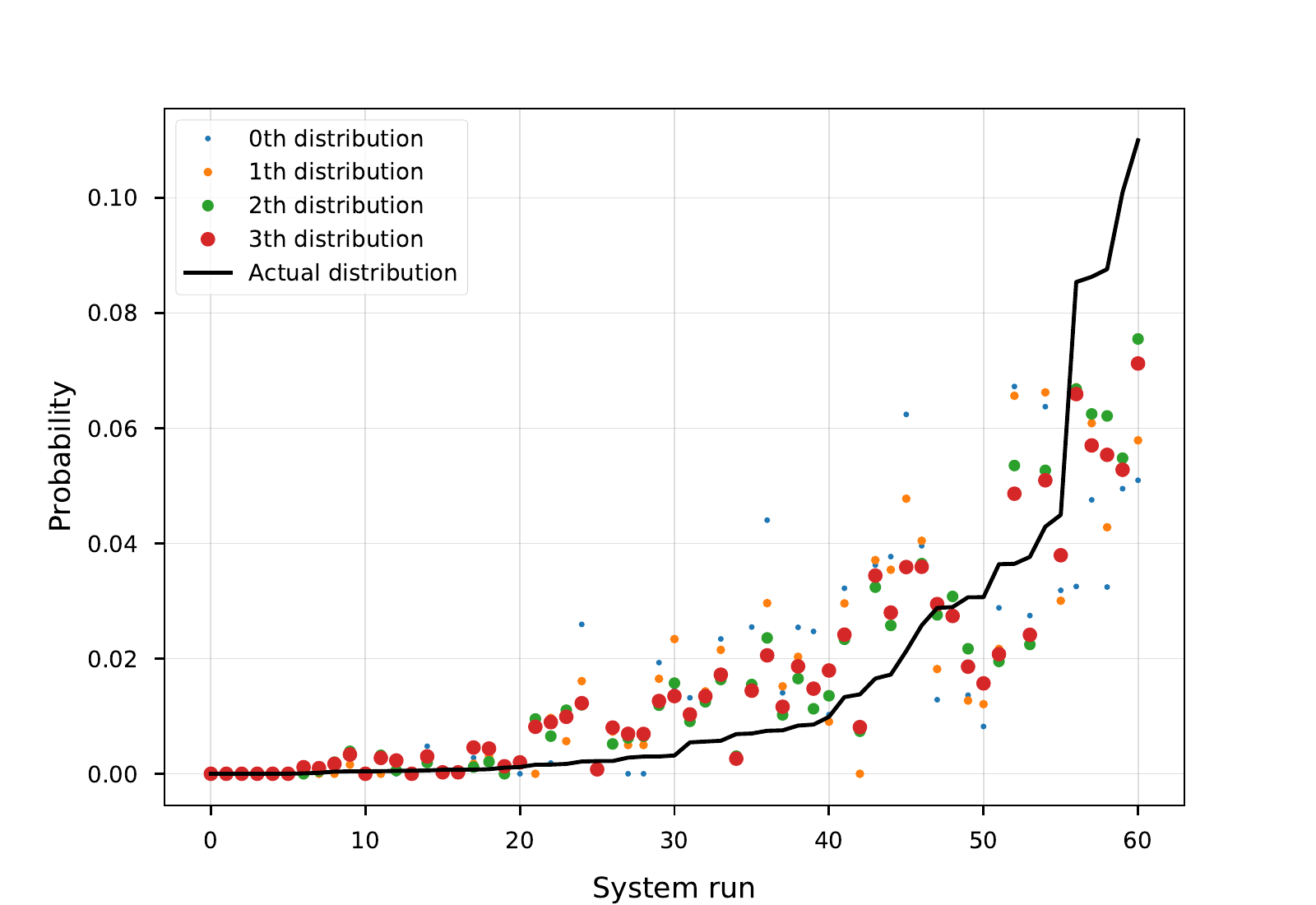}
\caption{Probability distribution over runs on topic 251 in TREC 5. The black curve is the probability induced by the actual average precisions based on depth-100 pooling, while circular markers of different sizes denote the approximate probabilities for different runs. Runs have been sorted according to their actual average precision values.}
\label{fig:SysrunDist}
\end{figure}

Any distribution that places a higher probability to better performing systems could be used here. In this work we consider the estimated performance of the retrieval systems on the basis of the relevance judgments accumulated at each round of assessments as system weights and normalize these weights to obtain a probability distribution over runs. Different evaluation measures can be used to estimate the performance of each run after every sampling round. 
Here we define a probability distribution proportional to estimated average precision $\widehat{AP}$ introduced in Section \ref{sec:evalmtrx}.
\begin{align*}
p_t(k) = \frac{\widehat{AP_t}(k)}{\sum_{k=1}^{K}{\widehat{AP_t}(k)}}, k=1,...,K; t=1,...,T
\end{align*}
Figure \ref{fig:SysrunDist} demonstrates the accuracy of the estimated (normalized) average precision at the end of four sampling rounds compared to the (normalized) average precision when the entire document collection (or to be more accurate the depth-100 pool for topic 251 in TREC 5) is used.  At every round the estimates (denoted with circular markers of different sizes for different rounds) better approximates the target values (denoted with a line). The details of the measure approximations are provided at Section~\ref{sec:estimator}.

\subsection{Distribution over document ranks}
The distribution over document ranks for a system $k$ determines the probability of sampling a document at a certain rank of the ordered list returned by run $k$. The underlying assumption that defines this probability distribution is that runs satisfy the Probability Ranking Principle (PRP)~\cite{Robertson:1997} which dictates that the probability of relevance monotonically decreases with the rank of the document. Hence, if we let $p$ denote the probability of sampling a document at rank $r$, then it is natural to assume $p$ is a function of $r$ and $p(r)$ monotonically decreases with $r$. Once again, any distribution that agrees with PRP can be used; researcher have used a number of such distributions (e.g. see ~\citet{aslam2005measure, pavlu2007practical, Hofmann:2011}).

In this work we consider an AP-prior distribution proposed by \citet{aslam2005measure} and \citet{pavlu2007practical} which aims to define the probability at each rank on the basis of the contribution of this rank in the calculation of average precision. The intuition is that when rewriting  $AP=\frac{1}{N}\sum_{1\leqslant j \leqslant i \leqslant N} \frac{1}{i} y_i y_j$, the implicit weight associated with rank $r$ can be obtained by summing weights associated with all pairs involving $r$, i.e. $\frac{1}{N}(1+\frac{1}{r}+\frac{1}{r+1}+...+\frac{1}{N})$. Then the AP-prior distribution is defined as follows:
\begin{align*}
&w(r)=\frac{1}{N}(1+\frac{1}{r}+\frac{1}{r+1}+...+\frac{1}{N}) \approx \frac{1}{N}\log{\frac{N}{r}} \\
&p(r)=\frac{w(r)}{\sum_{r=1}^{N}{w(r)}}
\end{align*}
where $r$ is the rank of a document and $N$ the total number of documents in the collection. Similar to \citet{aslam2005measure, pavlu2007practical} and all other sample-based methods, this distribution is defined at the beginning of the sampling process and remains fixed throughout the experiment.

%
%


\section{Retrieval performance estimator}
\label{sec:estimator}
In this section, we discuss the estimation of evaluation measures on the basis of the sampling procedure described in Algorithm~\ref{alg:sampling}. We first calculate the inclusion probabilities of each document in the collection, and then demonstrate how these probabilities can be used by a Horvitz-Thompson estimator to produce unbiased estimators of the population mean, and subsequently of some popular evaluation measures. The Horvitz-Thompson estimator, together with the calculated inclusion probabilities can be used to calculate the majority of the evaluation measures used in IR; in this paper we focus on three of them, Precision, Recall, and Average Precision. Other measures can be derived in similar ways (e.g. see Table 1 in \citet{Schnabel:2016}).


\subsection{Sampling with replacement with varying probabilities}


\textbf{Sampling procedure}. At each round of our iterative sampling process described in Algorithm~\ref{alg:sampling}, $n$ documents are sampled from a collection of size $N$. At each round, the unconditional probability of sampling a document $d_i$ (\textit{selection probability}) is $p_t(i)$, as defined in Step 2 of Algorithm~\ref{alg:sampling}, with 
\begin{eqnarray*}
  \sum_{i=1}^{N}{p_t(i)} = 1 & \textrm{and} & p_t(i) \ge 0 \\
  & \textrm{for} & i = 1,2,...,N\\
  & & t = 1, 2, ..., T.
\end{eqnarray*}
Let $i$ denote the index of the $n$ documents composing the sample set. The probability of a document $d_i$ being sampled (\textit{first-order inclusion probability}) at the end of the sampling process is given by
\[ 
\pi_i = 1 - \prod_{t=1}^{T}\prod_{z=1}^{N_t}{(1- p_t(i))}
\]
which accounts for varying probabilities across different rounds, while the probability of any two different document $d_i$ and $d_j$ being sampled (\textit{second-order inclusion probability}) is given by
\[
\pi_{ij} = \pi_i + \pi_j - [ 1 - \prod_{t=1}^{T}\prod_{z=1}^{N_t}{ (1- p_t(i) - p_t(j)) } ] 
\]
For the details of the derivation of the inclusion probabilities the reader can refer to \citet{booksampling}.
Using these inclusion probabilities together with the Horvitz-Thompson estimator allows us to construct unbiased estimators for different evaluation measures in IR.


\textbf{Horvitz-Thompson estimator of population total}.
\citet{horvitz1952generalization} propose a general sampling theory for constructing unbiased estimators of population totals. With any sampling design, with or without replacement, the unbiased Horvitz-Thompson estimator of the population total is 
\[
\widehat{\tau} = \sum_{i \in S'}{\frac{y_i}{\pi_i}}
\]
where $S'$ is the subset of $S$, only containing unique documents.  


An unbiased estimator of the variance of the population total estimator is given by:
\[
\widehat{var}(\mu) = \sum_{i \in S'}{(\frac{1}{{\pi_i}^2} - \frac{1}{\pi_i}){y_i}^2 } + 2 \sum_{i > j \in S'}{(\frac{1}{{\pi_i \pi_j}} - \frac{1}{\pi_{ij}})y_i y_j }
\]

For the details of these derivations the reader can refer to \citet{booksampling}.

\subsection{Evaluation metrics}
\label{sec:evalmtrx}
In this work we consider three of the most popular evaluation measures in IR, precision at a certain cut-off, PC(r), average precision, AP, and R-precision, RP.
We first clarify the exact expressions of the evaluation metrics with regard to the population, then introduce the estimators of these evaluation metrics on the sample set.
Let $C = \{d_i\}_{i=1}^{N}$ denote a population of  documents and let $y_i$ be an indicator variable of $d_i$, with $y_i = 1$ if the document $d_i$ is relevant, and $y_i = 0$ otherwise. The \textit{population total} is the summation of all $y_i$, i.e. the total number of relevant documents in the collection, while the \textit{population mean} is the population total divided by the \textit{population size}. If the population of documents considered is the documents ranked in the top-$r$ for some run $k$ then the population mean is the precision at cut-off $r$.

Based on the definition, \textit{precision at cutoff r}, \textit{average precision}, and \textit{precision at rank R} are defined as:
\begin{align*}
PC(r) = \frac{ \sum_{ d_i \in C, r(i) \leq r }{y_i} }{r} \\
AP = \frac{ \sum_{d_i \in C }{PC(r(i))y_i} }{R} \\
RP = \frac{ \sum_{ d_i \in C, r(i) \leq R }{y_i} }{R} \\
\end{align*}

Suppose that we have sampled $n$ documents $S=\{d_i\}_{i=1}^{n}$, with associated relevance labels $\{y_i\}_{i=1}^{n}$. We wish to estimate the total number of relevant documents in the collection, R, PC(r), AP and RP. Note that AP and RP, as many other evaluation measures that are normalized are ratios. For these measures, similar to previous work~\cite{pavlu2007practical} we can estimate the numerator and denominator separately, and while this ratio estimator is not guaranteed to be unbiased, the bias tends to be small and decrease with an increasing sample size~\cite{booksampling, raj1964note}.
%


The unbiased estimators for the four aforementioned measures based on Horvitz-Thompson can be calculated by:
\begin{align*}
&\widehat{R} = \sum_{d_i \in S'}{\frac{y_i}{\pi_i}} \\
&\widehat{PC}(r) =\frac{\sum_{d_i \in S', r(i) \leq r }{\frac{y_i}{\pi_i}} }{r} \\
&\widehat{AP} = \frac{\sum_{d_i \in S'}{\frac{\widehat{PC}(r(i))y_i}{\pi_i}}} {\widehat{R}} \\
&\widehat{RP} = \frac {\sum_{d_i \in S', r(i) \leq \widehat{R} }{\frac{y_i}{\pi_i}}} {\widehat{R}} \\
\end{align*}


%
%
%

\section{Experiment setup}

In this section we introduce our research questions, the statistics we use to evaluate the performance of the proposed estimators, and the data sets and baselines used in our experiments~\footnote{The implementation of the algorithm and the experiments run can be found at https://github.com/dli1/activesampling}. The batch size $N_b$ for all the experiments has been set to 3.

\subsection{Research questions}

In the remainder of the paper we aim to answer the following research questions:

\begin{description}
  \item[RQ1] How does active sampling perform compared to other sample-based and active-selection methods regarding bias and variance in the calculated effectiveness measures?
  \item[RQ2] How fast active sampling estimators approximate the actual evaluation measures compared to other sample-based and active-selection methods? 
  \item[RQ3] Is the test collection generated by active sampling reusable for new runs that do not contribute in the construction of the collection?
\end{description}

The aforementioned questions allow us to have a thorough examination of the effectiveness as well as the robustness of the proposed method.

\subsection{Statistics}
To answer the research questions put forward in the previous section, we need to quantify the performance of different document selection methods.

Our first goal is to measure how close the estimation of an evaluation measure is to its actual value when the full judgment set is being used. Assume that a document selection algorithm chooses a set of documents $S$ to calculate an evaluation measure. Let's denote the estimated measure with $f(k|S)$, for some run $k$. Let's also assume that the actual value of that evaluation measure, when the full judgment set is used, is $h(k)$. The \textit{root mean squared error (rms)}  of the estimator over  a sample set measures how close on average the estimated and the actual values are. We follow the definition in \cite{pavlu2007practical} : 
\[
 rms=\mathbb{E}_{S}\sqrt{\mathbb{E}_{k}{\big( f(k|S) - h(k) \big)}^2} . 
\] 
%

To further decompose the estimation errors made by different methods we also calculate the \textit{bias}, and the \textit{variance} decomposed from the \textit{mean square error (mse)} between the estimator and the corresponding real value. \textit{Bias} expresses the extent to which the average estimator over all sample sets differs from the actual value of a measure, while \textit{variance} expresses the extent to which the estimator is sensitive to the particular choice of a sample set (see \cite{bishop2006pattern}).  The \textit{mse}, $\mathbb{E}_{S}\mathbb{E}_{k}{\big( f(k|S) - h(k) \big)}^2$, can be rewritten as $\mathbb{E}_{k}\mathbb{E}_{S}{\big( f(k|S) - h(k) \big)}^2$, which can further be rewritten as ${ { {\Big ( \mathbb{E}_{S}{ \big( f(k|S) - h(k) \big)}\Big ) }^2 } } + { { \mathbb{VAR}_{S} f(k|S) } }$.
The first term denotes the \textit{bias} and second the \textit{variance} of the estimator. Taking all runs into account, we have
\begin{align*}
&bias = \mathbb{E}_{k} \mathbb{E}_{S}{ \big( f(k|S) - h(k) \big)}, \\
&variance = \mathbb{E}_{k} \mathbb{VAR}_{S} f(k|S). \\
\end{align*}
A second measurement we are interested in is how far the inferred ranking of systems when estimating an evaluation measure is to the actual ranking of systems when the entire judged collection is being used. Following previous work~\cite{Aslam:2003, aslam2006statistical, pavlu2007practical, Yilmaz:2006, Yilmaz:2008} we also report the \textit{Kendall's $\tau$} between estimated and actual rankings. Even though the \textit{Kendall's $\tau$} is an important measure when it comes to comparative evaluation, \textit{rms} error remains our focus, since test collections have found use not only in the evaluation of retrieval systems but also in learning retrieval functions~\cite{Li:2014}. In the latter case, for some algorithms, the accuracy of the estimated values is more important than just the correct ordering of systems.

\subsection{Test collections}

We conduct our experiments on TREC 5--8 AdHoc and TREC 9--11 Web tracks. The details of the data sets can be found in Table~\ref{tbl:TestCollection}. In our experiments we did not exclude any participating run, and we considered the relevance judgments released by NIST (qrels) as the complete set of judgment over which the actual values of measures are being computed. 

\begin{table*}[ht]
\centering
\caption{Test collections}
\label{tbl:TestCollection}
\begin{tabular}{c  c  c c c c c c}
\toprule
TREC       & Task type   & Topics    & \# runs & \# rel doc & \# judgement & \# rel doc per query & \# judgement per query  \\ \hline
TREC-5     & Adhoc        & 251-300 & 61                   & 5524 & 133681 & 110.48 & 2673.6    \\
TREC 6      & Adhoc       & 301-350 & 74                 & 4611 & 72270 & 92.22 & 1445.4   \\
TREC 7      & Adhoc      & 351-400  & 103              & 4674 & 80345 & 93.48 & 1606.9    \\
TREC 8      & Adhoc      & 401-450  & 129              & 4728 & 86830 & 94.56 & 1736.6   \\
TREC 9      & Web        & 451-500  & 104               & 2617 & 70070 & 52.34 & 1401.4    \\
TREC 10    & Web         & 501-550  & 97                 & 3363 & 70400 & 67.26 & 1408.0   \\
TREC 11    & Web         & 551-600   & 69               & 1574 & 56650 & 31.48 & 1133.0   \\
\bottomrule
\end{tabular}
\end{table*}

\subsection{Baselines}

We use two active-selection and one sample-based methods as baselines:

\noindent\textbf{Move-to-Front (MTF)} \cite{cormack1998efficient}.
MTF is a deterministic, iterative selection method. At the first round, all runs are given equal priorities. At each round, the method selects the run with the highest priority and obtains the judgment of the first unjudged document in the ranked list of the given run. If the document is relevant the method selects the next unjudged document until a non-relevant document is judged. If that happens the priority of the current run is being reduced and the run with the highest priority is selected next.

\noindent\textbf{Multi-armed Bandits (MAB)} \cite{losada2016feeling}.
Similar to MTF, MAB aims to find as many relevant documents as possible. MAB casts document selection as a multi-armed bandit problem, and different to MTF it randomly decides whether to select documents from the best run on the current stage, or sample a document across the entire collection. For the MAB baseline we used the best method \textbf{MM-NS} with its default setting reported in \cite{losada2016feeling}~\footnote{http://tec.citius.usc.es/ir/code/pooling_bandits.html}. 

\noindent\textbf{Stratified sampling} \cite{pavlu2007practical}.
Stratified sampling is a stochastic method based on \textit{importance sampling}. The probability distribution over documents used is the AP-prior distribution, which remains unchanged throughout the sampling process. Similar to our approach, the Horvitz-Thompson estimator is used to estimate the evaluation metrics. The stratified sampling approach proposed by \citet{pavlu2007practical} has been used in the construction of the TREC Million Query track collection~\cite{Carterette:2009}, it outperforms methods using uniform random sampling~\cite{aslam2006statistical, Yilmaz:2006} and demonstrate similar performance to ~\citet{Yilmaz:2008}.
 

\section{Results and analysis}

\subsection{Bias and Variance}
\label{section:rq1}

\begin{figure*}[t]
\centering
\includegraphics[width = 1\textwidth]{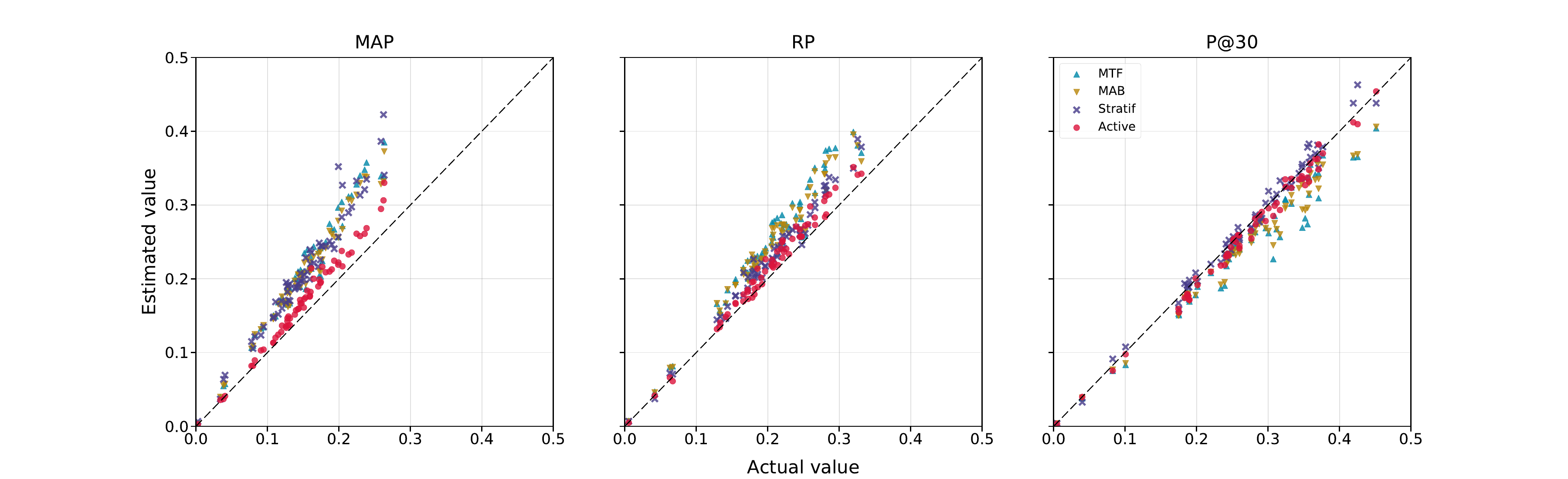}
\caption{Estimated vs actual values of MAP, RP, and PC@30 for different runs on a 10\% sample of TREC 5.}
\label{fig:Exp1-1}
\end{figure*}

\begin{figure*}[t]
\centering
\includegraphics[width = 1\textwidth]{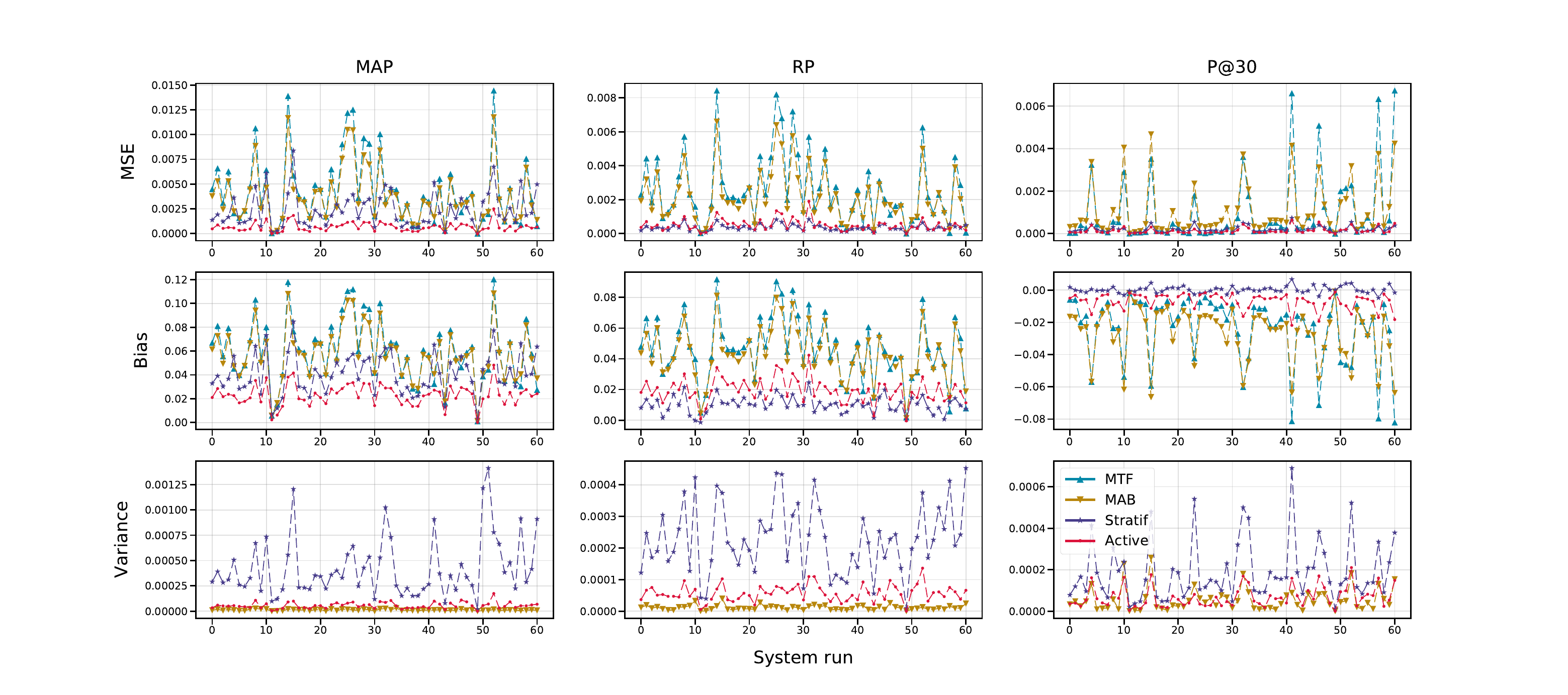}
\caption{RMS error , variance, and bias (y-axis) of the sample-based and active-selection methods compared, for different runs (x-axis) over 30 sample sets drawn from TREC 5. The \textit{mse} of Active is significantly smaller than that of MTF/MAB/Stratif for MAP/RP/P@30 at 95\% confidence level by Welch's t-test, except for Stratif on RP where Active is significantly larger than Stratif. }
\label{fig:Exp1-2}
\end{figure*}

This first experiment is designed to answer RQ1 and is conducted on TREC 5. 

We reduce the retrieved document lists of all runs to the top-100 ranks (so that all documents in the ranked lists are judged) and consider this the ground truth rankings, based on which the actual values of MAP, RP and P@30 are calculated. The judgment effort is set to 10\% of the depth-100 pool for each query, and different methods are used to obtain the corresponding subset and calculate the estimated MAP, RP and P@30 for each run. For any stochastic method (i.e. the sampling methods and MAB) the experiment is repeated 30 times. Based on the estimated and actual values we calculate $mse$, and its decomposition to $bias$ and $variance$ for each estimator.

Figure \ref{fig:Exp1-1} shows a number of scatter plots for MTF, MAB, Stratif (stratified sampling), and our method denoted as Active (active sampling). Each point in the plots corresponds to a given run. To declutter the figure, the shown points for the sample-based methods are computed over a single sample. An unbiased estimator should lead to points that lie on the x=y line. As it can be observed the active sampling estimated values are the ones that are closer to the diagonal. As expected, and by construction, precision is unbiased, while the bias introduced in the ratio estimators of AP and RP is smaller that all active-selection methods, and comparable to the stratified sampling method. 

A decomposition of the mse into bias and variance can be found in Figure \ref{fig:Exp1-2}. 
As expected the variance of active-selection method is zero (or close to zero) since MTF is a deterministic method, while the randomness of MAB is only in the decision between exploration and exploitation. Active sampling has a much lower variance than stratified sampling, which demonstrates one of the main contribution of the our sampling method: biasing the sampling distribution towards good performing runs improves the estimation of the evaluation measures. The bias of the sample-based methods, as expected, is near-zero, while it is smaller than zero for the active-selection methods, since they do not correct for their preference to select documents from good performing runs. For example, the bias on P@30 of active-selection methods are much smaller than zero, because the greedy strategies only count the number of relevant documents and thus underestimate P@30; while the sampling methods can avoid the problem by using unbiased estimators. This demonstrates the second main contribution of our approach: using sampling avoids any systematic error in the calculation of measures. Therefore, the proposed sampling method indeed combines the advantages of both sample-based and active-selection methods that have been proposed in the literature.

%

\subsection{Effectiveness}
\label{section:rq2}

\begin{figure*}[t]
  \centering
  \includegraphics[width = 1\textwidth]{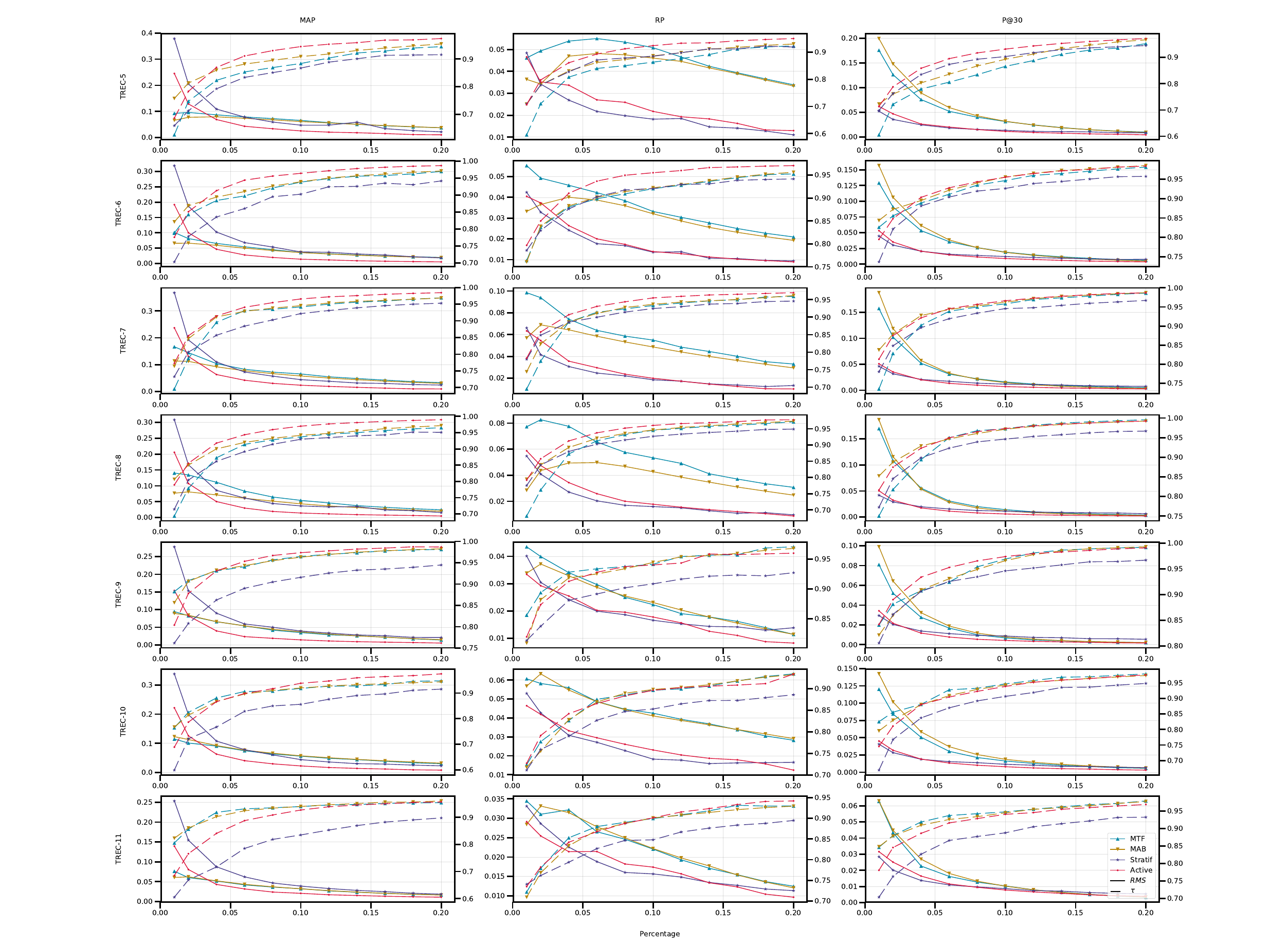}
\caption{Average $rms$ and $\tau$ for MTF, MAB, Stratified and Active sampling at different sample sizes in TREC 5-11. The left y-axis and solid lines denote $rms$, the right y-axis and dotted lines denote Kendall's $\tau$. }
\label{fig:Exp2-1}
\end{figure*}

This second experiment is designed to answer RQ2 and is conducted on TREC 5-11. In this experiment we vary the judgment effort from 1\% to 20\% of the depth-100 pool. At each sampling percentage, when sample-based methods are used, we first calculate the $rms$ error and Kendall's $\tau$ values for a given sample and then average these values over 30 sample sets.

\begin{figure}[b]
\centering
\includegraphics[width = 1\columnwidth]{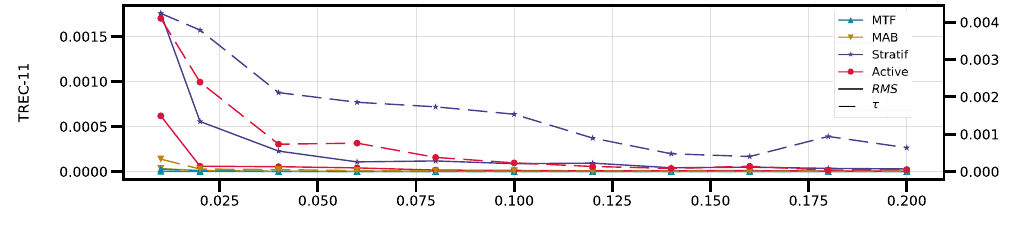}
\caption{Variance of rms error (solid lines) and Kendall's $\tau$ values (dashed lines) when estimating MAP, over 30 sample for different sample sizes TREC 11.}
\label{fig:Exp2-2}
\end{figure}

Figure \ref{fig:Exp2-1} shows the average $rms$ and $\tau$ value at different sample sizes. For all TREC tracks active sampling demonstrates a lower rms error than stratified sampling, MTF, and MAB for sampling rates greater than 3-5\%. At lower sampling rates active-selection methods show an advantage compared to sample-based methods that suffer from high variance. Regarding Kendall's $\tau$ active sampling outperforms all methods for TREC 5--8, for sampling rates greater than 5\%, while for TREC 10 and 11 it picks up at sampling rates greater than 10\%. TREC 10 and 11 are the two collection with the smallest number of relevant documents per query, hence finding these document using active-selection methods leads to a better ordering of systems when the percentage of judged documents is very small. For those small percentages the sample-based methods demonstrate high variance, and it really depends on how lucky one is when drawing the sample of documents. The variance of $rms$ error and Kendall's $\tau$ across the 30 different samples drawn in this experiment for the estimation of MAP on TREC 11 can be seen in Figure~\ref{fig:Exp2-2}.

Overall, when comparing active sampling with MTF and MAB, we find that our method outperforms them regarding $rms$. This indicates once again that the calculated inclusion probabilities and the Horvitz-Thompson estimator allows active sampling to produce an unbiased estimation of the actual value of the evaluation measures. When comparing active sampling with stratified sampling, both of which use the Horvitz-Thompson estimator, we can find that our method outperforms stratified sampling regarding Kendall's $\tau$. This indicates that the dynamic strategy we employ is beneficial compared to a static sampling distribution. Therefore, active sampling indeed combines the advantage of both methods.

\subsection{Reusability}
\begin{figure*}[t]
\centering
\includegraphics[width = 1\textwidth]{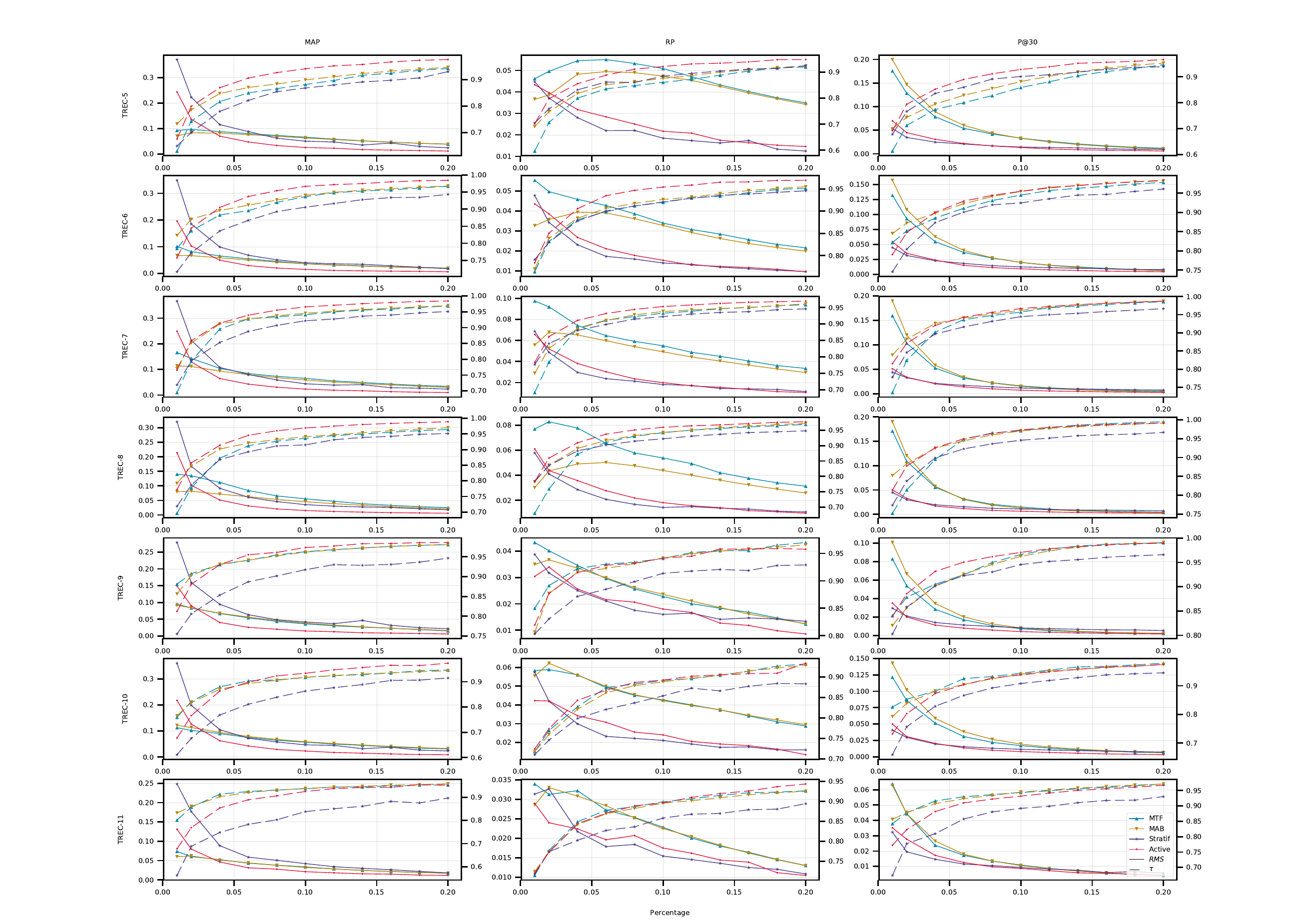}
\caption{Average $rms$ and $\tau$ of MTF, MAB, Stratified and Active sampling at different sample sizes on leave-one-group-out runs in TREC 5-11.}
\label{fig:Exp3}
\end{figure*}

Constructing a test collection is a laborious task, hence it is very important that the proposed document selection methods construct test collections that can be used to evaluate new, novel algorithms without introducing any systematic errors.
This experiment is designed to answer RQ3 and is conducted on TREC 5-11. In this experiment we split the runs into contributing runs and left-out runs. Using the contributing runs we construct a test collection for each different document selection method. We then calculate the estimated measures for all runs including those that were left out from the collection construction experiment.
In our experiment, we use a one-group-out split of the runs. Runs that contributed in the sampling procedure come from different participating groups. Groups often submit different versions of the same retrieval algorithm, hence, typically, all the runs submitted by the same participating group differ very little in the ranking of the documents. To ensure that left-out runs are as novel as possible we leave out all runs for a given group.
Regarding the calculation of rms error and Kendall's $\tau$ we compute rms error and Kendall's $\tau$ considering both participating and left-out runs. 

Figure \ref{fig:Exp3} shows the average $rms$ error and Kendall's $\tau$ values at different sample sizes using the latter afore-described option to isolate the effect of the different document selection methods on new, novel systems.
In general, the trends observed in Figure~\ref{fig:Exp2-1} can also be observed in Figure~\ref{fig:Exp3}, with active sampling outperforming all other methods regarding rms error and Kendall's $\tau$ for sampling rates greater than 5\%. For sampling rates lower than 5\% in collections with very few relevant documents per topic (such as TREC 10 and 11) the active-selection methods perform better than the sample-based methods, however we can also conclude that at these low sampling rates none of the methods lead to reliably reusable collections.


\section{Conclusion}

In this paper we consider the problem of large-scale retrieval evaluation. We tackle the problem of devising a sample-based approach - \textit{active sampling}. Our method consists of a sampling step and an unbiased estimation step. In the sampling step, we construct two distributions, one over retrieval systems that is updated at every round of relevance judgments giving larger probabilities to better quality runs, and one over document ranks that is defined at the beginning of the sampling process and remains static throughout the experiment. Document samples are drawn from the joint probability distribution, and inclusion probabilities are computed at the end of the entire sampling process accounting for varying probabilities across sampling rounds. In the estimation step, we use the well-known Horvitz-Thompson estimator to estimate evaluation metrics for all system runs.

The proposed method is designed to combine the advantages of two different families of methods that have appeared in the literature: sample-based and active-selection approaches. Similar to the former, our method leads to unbiased, by construction, estimators of evaluation measures, and can safely be used to evaluate new, novel runs that have not contributed to the generation of the test collection. Similar to the latter, the attention of our method is put on good quality runs with the hope of identifying more relevant documents and reduce the variability naturally introduced in the estimation of a measure due to sampling.

To examine the performance of the proposed method, we tested against state-of-the-art sample-based and active-selection methods over seven TREC AdHoc and Web collections, TREC 5--11. Compared to sample-based approaches, such as stratified sampling, out method indeed demonstrated lower variance, while compared against active-selection approaches, such as Move-to-Front, and Multi-Armed Bandits, our method, as expected, has lower, near-zero bias. For sampling rates as low as 5\% of the entire depth-100 pool, the proposed method outperforms all other methods regarding effectiveness and efficiency and leads to reusable test collections.


\begin{acks}
This research was supported by the Google Faculty Research Award program. All content represents the opinion of the authors, which is not necessarily shared or endorsed by their respective employers and/or sponsors. 
\end{acks}

\bibliographystyle{ACM-Reference-Format}
\bibliography{references} 
\end{document}